\documentclass[aps,prd,preprint,groupedaddress,showpacs]{revtex4}

\usepackage{epsfig}
\usepackage{graphics}

\begin{document}

\def\choosen{\atopwithdelims..}

\preprint{DESY~06--104\hspace{11.5cm} ISSN 0418-9833}
\preprint{July 2006\hspace{14.9cm}}

\boldmath
\title{Bottomonium Production in the Regge Limit of QCD}
\unboldmath

\author{\firstname{B.A.} \surname{Kniehl}}
\email{kniehl@desy.de}
\affiliation{{II.} Institut f\"ur Theoretische Physik, Universit\" at Hamburg,
Luruper Chaussee 149, 22761 Hamburg, Germany}

\author{\firstname{V.A.} \surname{Saleev}}
\email{saleev@ssu.samara.ru}
\author{\firstname{D.V.} \surname{Vasin}}
\email{vasin@ssu.samara.ru}
\affiliation{Department of Physics, Samara State University,
Academic Pavlov Street~1, 443011 Samara, Russia}

\begin{abstract}
We study bottomonium hadroproduction in proton-antiproton collisions at the
Fermilab Tevatron in the framework of the quasi-multi-Regge kinematics
approach and the factorization formalism of non-relativistic QCD at leading
order in the strong-coupling constant $\alpha_s$ and the relative velocity $v$
of the bound quarks.
The transverse-momentum distributions of prompt $\Upsilon(nS)$-meson
production measured at the Tevatron are fitted to obtain the non-perturbative
long-distance matrix elements for different choices of un-integrated gluon
distribution functions of the proton.
\end{abstract}


\maketitle \maketitle

\section{Introduction}

Bottomonium production at high energies has provided a useful laboratory for
testing the high-energy limit of quantum chromodynamics (QCD) as well as the
interplay of perturbative and non-perturbative phenomena in QCD.
The factorization formalism of non-relativistic QCD (NRQCD) \cite{NRQCD} is a
rigorous theoretical framework for the description of heavy-quarkonium
production and decay.
The factorization hypothesis of NRQCD assumes the separation of the effects of
long and short distances in heavy-quarkonium production.
NRQCD is organized as a perturbative expansion in two small parameters, the
strong-coupling constant $\alpha_s$ and the relative velocity $v$ of the
heavy quarks.

The phenomenology of strong interactions at high energies exhibits a dominant
role of gluon interactions in quarkonium production.
In the conventional parton model \cite{PartonModel}, the initial-state gluon
dynamics is controlled by the Dokshitzer-Gribov-Lipatov-Altarelli-Parisi
(DGLAP) evolution equation \cite{DGLAP}.
In this approach, it is assumed that $S > \mu^2 \gg \Lambda_{\rm QCD}^2$,
where $\sqrt{S}$ is the invariant collision energy, $\mu$ is the typical
energy scale of the hard interaction, and $\Lambda_{\rm QCD}$ is the
asymptotic scale parameter of QCD.
In this way, the DGLAP evolution equation takes into account only one big
logarithm, namely $\ln(\mu/\Lambda_{\rm QCD})$.
In fact, the collinear-partron approximation is used, and the transverse
momenta of the incoming gluons are neglected.

In the high-energy limit, the contribution from the partonic subprocesses
involving $t$-channel gluon exchanges to the total cross section can become
dominant.
The summation of the large logarithms $\ln(\sqrt{S}/\mu)$ in the evolution
equation can then be more important than the one of the
$\ln(\mu/\Lambda_{\rm QCD})$ terms.
In this case, the non-collinear gluon dynamics is described by the
Balitsky-Fadin-Kuraev-Lipatov (BFKL) evolution equation \cite{BFKL}.
In the region under consideration, the transverse momenta ($k_T$) of the
incoming gluons and their off-shell properties can no longer be neglected,
and we deal with reggeized $t$-channel gluons.
The theoretical frameworks for this kind of high-energy phenomenology are the
$k_T$-factorization approach \cite{KTGribov,KTCollins} and the
quasi-multi-Regge kinematics (QMRK) approach
\cite{KTLipatovFadin,KTAntonov}, which is based on effective quantum field
theory implemented with the non-abelian gauge-invariant action, as suggested
a few years ago \cite{KTLipatov}.
Our previous analysis of charmonium production at high-energy colliders using
the high-energy factorization scheme \cite{KniehlSaleevVasin,PRD2003} has
shown the equivalence of the $k_T$-factorization and the QMRK approaches at
leading order (LO) in $\alpha_s$.
However, the $k_T$-factorization approach has well-known principal
difficulties \cite{smallx} at next-to-leading order (NLO).
By contrast, the QMRK approach offers a conceptual solution of the NLO
problems \cite{Ostrovsky}.
In our LO applications, the QMRK approach yields similar formulas as the
$k_T$-factoriztion approach, so that we can essentially continue using our
previous results \cite{KniehlSaleevVasin,PRD2003} obtained in the
$k_T$-factorization formalism using the Collins-Ellis prescription
\cite{KTCollins}.

This paper is organized as follows.
In Sec.~\ref{sec:two}, the QMRK approach is briefly reviewed.
In Sec.~\ref{sec:four}, we explain how the analytic results of
Refs.~\cite{KniehlSaleevVasin,PRD2003} relevant for our analysis may be
converted to the QMRK framework.
In Sec.~\ref{sec:five}, we perform fits to the transverse-momentum
($p_T=|{\bf p}_T|$) distributions of inclusive bottomonium production measured
at the Fermilab Tevatron to obtain numerical values for the non-perturbative
matrix elements (NMEs) of the NRQCD factorization formalism.
In Sec.~\ref{sec:six}, we summarize our results.

\boldmath
\section{QMRK approach}
\unboldmath
\label{sec:two}

In the phenomenology of strong interactions at high energies, it is necessary
to describe the QCD evolution of the gluon distribution functions of the
colliding particles starting from some scale $\mu_0$, which controls a
non-perturbative regime, to the typical scale $\mu$ of the hard-scattering
processes, which is typically of the order of the transverse mass
$M_T=\sqrt{M^2+|{\bf p}_T|^2}$ of the produced particle (or hadron jet) with
(invariant) mass $M$ and transverse two-momentum ${\bf p}_T$.
In the region of very high energies, in the so-called Regge limit, the typical
ratio $x=\mu/\sqrt{S}$ becomes very small, $x\ll1$.
This leads to large logarithmic contributions of the type
$[\alpha_s\ln(1/x)]^n$ in the resummation procedure, which is described by the
BFKL evolution equation \cite{BFKL} for an un-integrated gluon distribution
function $\Phi(x,|{\bf q}_T|^2,\mu^2)$, where ${\bf q}_T$ is the transverse
two-momentum of the gluon with respect to the flight direction of the incoming
hadron from which it stems.
Accordingly, in the QMRK approach \cite{KTLipatovFadin}, the initial-state
$t$-channel gluons are considered as reggeons (or reggeized gluons).
They carry finite transverse two-momenta ${\bf q}_T$ with respect to the
hadron beam from which they stem and are off mass shell.

\begin{figure}[hpt]
\begin{center}
\psfig{figure=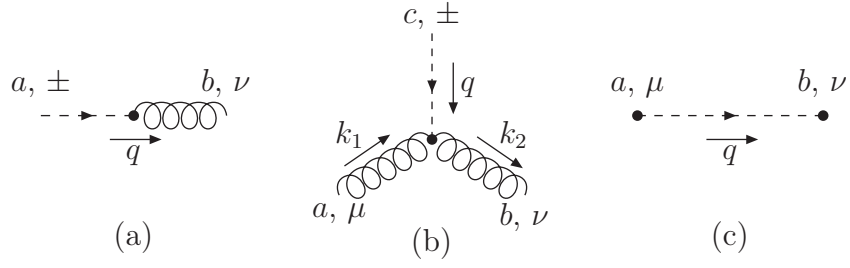}
\caption{\label{fig:FR}Feynman diagrams pertinent to the (a) PR vertex, (b)
PPR vertex, and (c) reggeized-gluon propagator given in Eqs.~(\ref{eq:PR}),
(\ref{eq:PPR}), and (\ref{eq:RR}), respectively.}
\end{center}
\end{figure}

Reggeized gluons interact with quarks and Yang-Mills gluons in a specific way.
Recently, in Ref.~\cite{KTAntonov}, the Feynman rules for the effective theory
based on the non-abelian gauge-invariant action \cite{KTLipatov} were derived
for the induced and some important effective vertices.
The induced vertex for the transition from a reggeized gluon to a Yang-Mills
gluon $R^{\pm}\to g $ (PR vertex) shown in Fig.~\ref{fig:FR}(a) has the form:
\begin{equation}
\Gamma_{ab}^{\pm\nu}(q) = i \delta^{ab} q^2 (n^{\pm})^\nu,
\label{eq:PR}
\end{equation}
where $(n^+)^\nu = P_1^\nu/E_1$,
$(n^-)^\nu = P_2^\nu/E_2$,
$P_{1,2}^\nu$ are the four-momenta of the colliding protons, and
$E_{1,2}$ are their energies in the center-of-mass frame.
We have $(n^\pm)^2=0$, $n^+\cdot n^-=2$, and $S=(P_1+P_2)^2=4E_1E_2$.
For any four-momentum $k^{\mu}$, we define $k^{\pm}=k\cdot n^{\pm}$.
It is easy to see that the four-momenta of the reggeized gluons can be
represented as
\begin{eqnarray}
q_1^\mu&=&q_{1T}^\mu + \frac{q_1^-}{2}(n^+)^\mu,
\nonumber\\
q_2^\mu&=& q_{2T}^\mu + \frac{q_2^+}{2}(n^-)^\mu,
\nonumber\\
q_1^+&=&q_2^- = 0.
\end{eqnarray}
The induced interaction vertex of one reggeized gluon with two Yang-Mills
gluons (PPR vertex) depicted in Fig.~\ref{fig:FR}(b) reads
\begin{equation}
\Gamma^{\mu\pm\nu}_{acb}(k_1,q,k_2) = - g_s f^{abc}
\frac{q^2}{k_1^\pm}(n^{\pm})^\mu (n^{\pm})^\nu,
\label{eq:PPR}
\end{equation}
where $g_s=\sqrt{4\pi\alpha_s}$ is the gauge coupling of QCD.
The reggeized-gluon propagator displayed in Fig.~\ref{fig:FR}(c) is given by
\begin{equation}
D^{\mu\nu}_{ab}(q) = -i \delta^{ab}\frac{1}{2 q^2}
\left[(n^+)^\mu(n^-)^\nu + (n^+)^\nu(n^-)^\mu \right].
\label{eq:RR}
\end{equation}
The Lagrangian of the effective theory~\cite{KTLipatov} also contains the
standard gluon-gluon and quark-gluon interactions for Yang-Mills gluons.

\begin{figure}[hpt]
\begin{center}
\psfig{figure=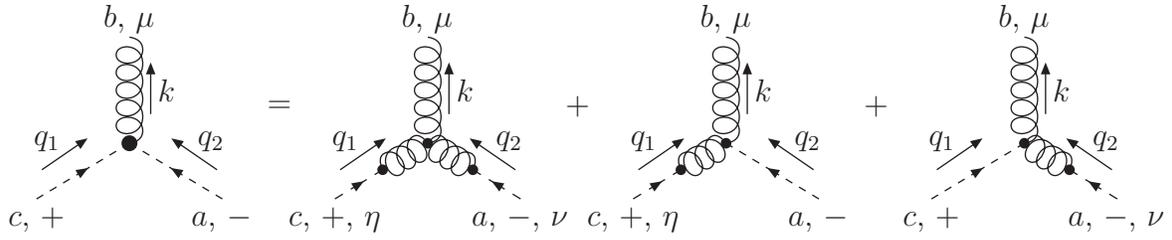}
\caption{\label{fig:EffectiveVertex}Feynman diagrams pertinent to the
effective PRR vertex given in Eq.~(\ref{eq:PRR}).}
\end{center}
\end{figure}

Using the Feynman rules for the induced vertices (\ref{eq:PR}) and
(\ref{eq:PPR}) and the ordinary vertices, one can construct effective
vertices, which obey Bose and gauge symmetries.
For example, the effective three-vertex that describes the production of a
single Yang-Mills gluon with four-momentum $k^\mu=q_1^\mu+q_2^\mu$ and color
index $b$ by two-reggeon annihilation $R^++R^-\to g$ (PRR vertex) shown in
Fig.~\ref{fig:EffectiveVertex} reads
\begin{eqnarray}
\Gamma^{+\mu-}_{cba}(q_1,k,q_2) &=&
V^{\rho\sigma\mu}_{cab}(-q_1,-q_2,k) (n^+)_\rho (n^-)_\sigma
+\Gamma^{\rho-\mu}_{cab}(q_1,q_2,k) (n^+)_\rho
+\Gamma^{\sigma+\mu}_{acb}(q_2,q_1,k) (n^-)_\sigma
\nonumber\\
&=&2g_sf^{cba}\left[
(n^-)^\mu\left(q_2^++\frac{q_2^2}{q_1^-}\right)
-(n^+)^\mu\left(q_1^-+\frac{q_1^2}{q_2^+}\right)+(q_1-q_2)^\mu  \right],
\label{eq:PRR}
\end{eqnarray}
where
\begin{equation}
V^{\lambda\mu\nu}_{abc}(k_1,k_2,k_3) = - g_s f^{abc}
\left[(k_1-k_2)^\nu g^{\lambda\mu}+(k_2-k_3)^\lambda g^{\mu\nu}
+(k_3-k_1)^\mu g^{\nu\lambda}\right]
\end{equation}
is the Yang-Mills three-gluon vertex, with all four-momenta taken to be
outgoing, and we exploited the following relation
\begin{equation}
\delta^{ab}(n^{\pm})^\mu=\Gamma^{\pm\nu}_{ac}(q)
\left(-i\delta^{cb}\frac{g^{\mu\nu}}{q^2}\right).
\end{equation}

The gauge invariance of the effective theory \cite{KTLipatov} leads to the
following condition for amplitudes in the QMRK:
\begin{equation}
\lim_{|{\bf q}_{1T}|,|{\bf q}_{1T}|\to 0} \overline{|{\cal A}(R+R
\to {\cal H}+X)|^2}=0.
\end{equation}

In the QMRK approach, the hadronic cross section of quarkonium (${\cal H}$)
production through the process
\begin{equation}
p + \bar p \to {\cal H} + X
\label{eq:ppHX}
\end{equation}
and the partonic cross section of the two-reggeon fusion subprocess
\begin{equation}
R + R \to {\cal H} + X
\label{eq:RRHX}
\end{equation}
are related as
\begin{eqnarray}
d\sigma(p + \bar p \to {\cal H} + X)&=&  \int\frac{d x_1}{x_1}
\int\frac{d^2{\bf q}_{1T}}{\pi} \Phi\left(x_1,|{\bf q}_{1T}|^2,\mu^2\right)
\int\frac{d x_2}{x_2} \int\frac{d^2 {\bf q}_{2T}}{\pi}
\nonumber\\
&&{}\times\Phi\left(x_2,|{\bf q}_{2T}|^2,\mu^2\right)
d\hat\sigma(R + R \to {\cal H}+X).
\label{eq:KT}
\end{eqnarray}
where $\Phi\left(x,|{\bf q}_{T}|^2,\mu^2\right)$ is the un-integrated gluon
distribution function in the proton,
$x_1=q_1^-/(2E_1)$ and $x_2 =q_2^+/(2 E_2)$ are the fractions of the proton
momenta passed on to the reggeized gluons, and the factorization scale $\mu$
is chosen to be of order $M_T$.
The collinear and un-integrated gluon distribution functions are formally
related as
\begin{equation}
xG(x,\mu^2)=\int_0^{\mu^2} d|{\bf q}_{T}|^2
\Phi\left(x,|{\bf q}_{T}|^2,\mu^2\right),
\label{eq:xG}
\end{equation}
so that, for ${\bf q}_{1T}={\bf q}_{2T}={\bf 0}$, we recover the conventional
factorization formula of the collinear parton model,
\begin{equation}
d\sigma(p + \bar p \to {\cal H}+\!X)=\int{d x_1}G(x_1,\mu^2)
\int{d x_2} G(x_2,\mu^2) d\hat \sigma(g + g \to {\cal H} + X).
\label{eq:PM}
\end{equation}

The partonic cross section or process~(\ref{eq:RRHX}) may be evaluated as
\begin{equation}
d\hat\sigma(R + R\to {\cal H}+X)
=\frac{{\cal N}}{2x_1x_2S}\overline{|{\cal A}(R + R\to {\cal H}+X)|^2}d\Phi,
\end{equation}
where $2x_1x_2S$ is the flux factor of the incoming particles,
${\cal A}(R + R\to {\cal H}+X)$ is the production amplitude, the
overbar indicates average (summation) over initial-state
(final-state) spins and colors, $d\Phi$ is the phase space volume of
the outgoing particles, and
\begin{equation}
{\cal N}= \frac{(x_1x_2S)^2}{16|{\bf q}_{1T}|^2|{\bf q}_{2T}|^2}
\label{eq:Norm}
\end{equation}
is a normalization factor that ensures the correct transition to the
collinear-parton limit.
This convention implies that the partonic cross section in the QMRK approach
is normalized approximately to the cross section for on-shell gluons when
${\bf q}_{1T}={\bf q}_{2T}={\bf0}$.

In our numerical calculations, we use the un-integrated gluon distribution
functions by Bl\"umlein (JB) \cite{JB}, by Jung and Salam (JS) \cite{JS}, and
by Kimber, Martin, and Ryskin (KMR) \cite{KMR}.
A direct comparison between different un-integrated gluon distributions as
functions of $x$, $|{\bf k}_T|^2$, and $\mu^2$ may be found in
Ref.~\cite{PLB2002}.
Note that the JB version is based on the BFKL evolution equation \cite{BFKL}.
On the contrary, the JS and KMR versions were obtained using the more
complicated Catani-Ciafaloni-Fiorani-Marchesini (CCFM) evolution equation
\cite{CCFM}, which takes into account both large logarithms of the types
$\ln(1/x)$ and $\ln(\mu/\Lambda_{\rm QCD})$.

\boldmath
\section{Relation between QMRK and
$k_T$-factorization approaches}
\label{sec:four}
\unboldmath

In this section, we obtain the squared amplitudes for inclusive quarkonium
production via the fusion of two reggeized gluons in the framework of
QMRK~\cite{KTAntonov} and NRQCD~\cite{NRQCD}.
We work at LO in $\alpha_s$ and $v$ and consider the following partonic
subprocesses \cite{KniehlSaleevVasin}:
\begin{eqnarray}
R + R &\to& {\cal H} \left[{^3P}_J^{(1)},{^3S}_1^{(8)},{^1S}_0^{(8)},
{^3P}_J^{(8)}\right],
\label{eq:RRtoH}\\
R + R &\to& {\cal H} \left[{^3S}_1^{(1)}\right] + g.
\label{eq:RRtoHG}
\end{eqnarray}
This formalism also allows for a consistent treatment at NLO,  which is,
however, beyond the scope of this paper and needs a separate investigation.

\begin{figure}[hpt]
\begin{center}
\psfig{figure=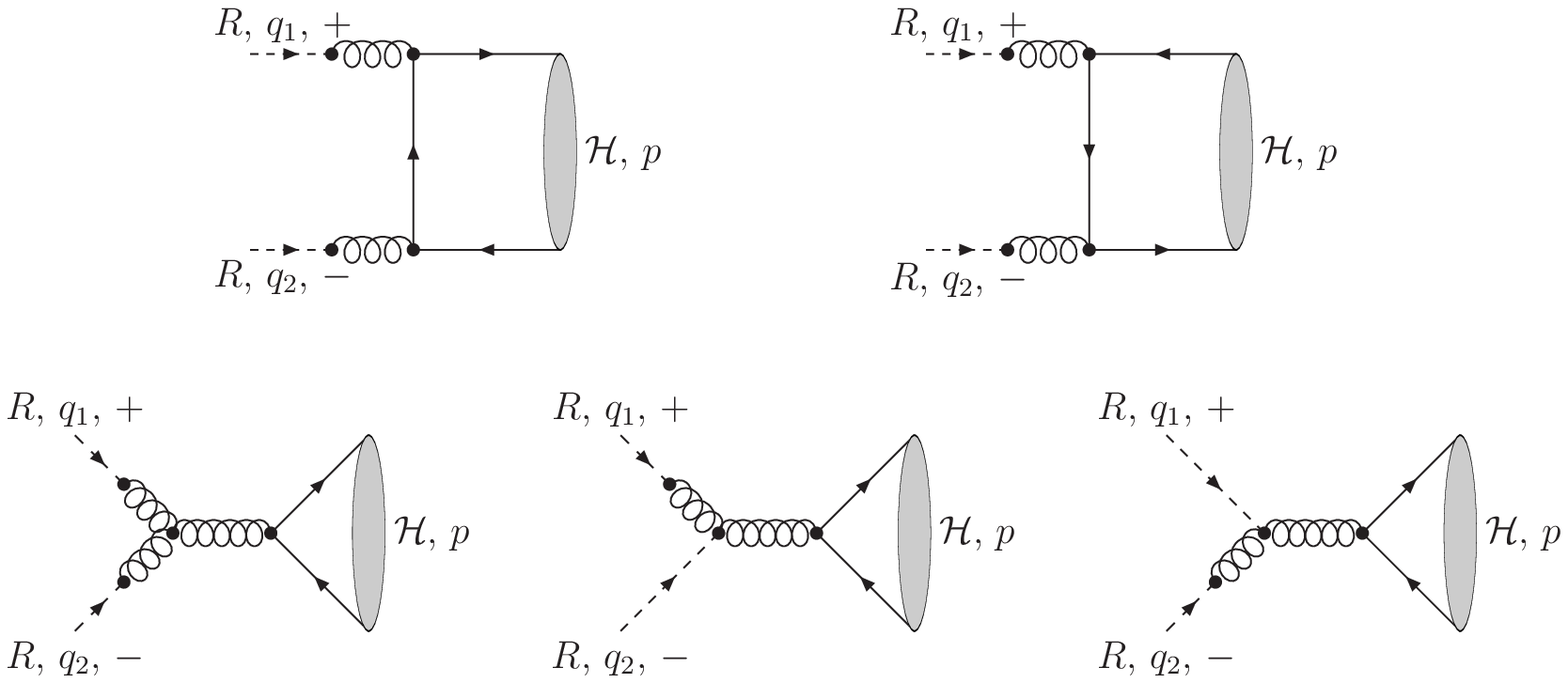,width=1.0\textwidth}
\caption{\label{fig:RRH}Feynman diagrams pertinent to
processes~(\ref{eq:RRtoH}).}
\end{center}
\end{figure}

\begin{figure}[ht]
\begin{center}
\psfig{figure=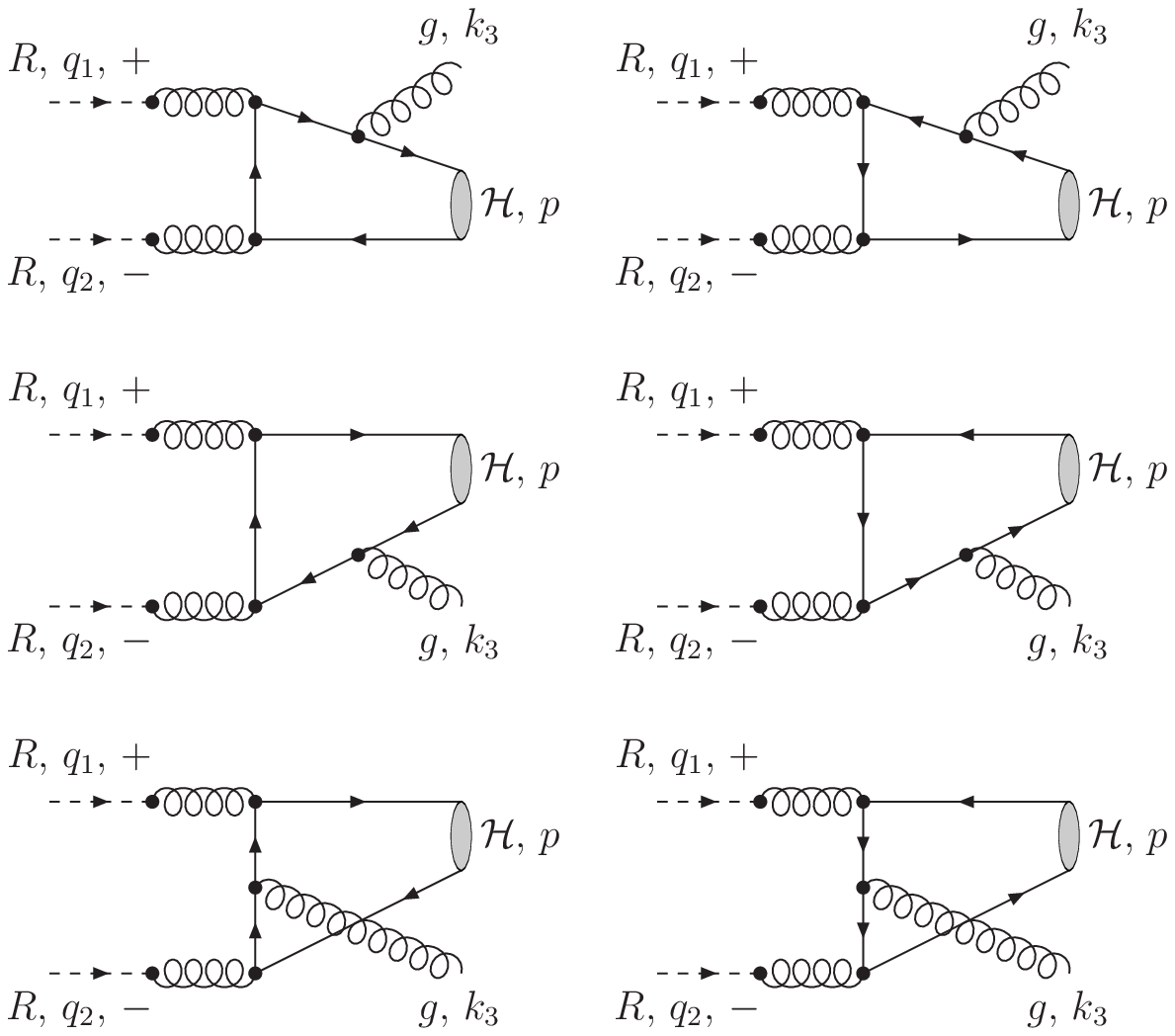,width=0.75\textwidth}
\caption{\label{fig:RRHg}Feynman diagrams pertinent to
process~(\ref{eq:RRtoHG}).}
\end{center}
\end{figure}

According to the prescription of Ref.~\cite{KTAntonov}, the amplitudes of
processes (\ref{eq:RRtoH}) may be obtained from the five Feynman diagrams
depicted in Fig.~\ref{fig:RRH}.
Of course, the last three Feynman diagrams in Fig.~\ref{fig:RRH} can be
combined through the effective PRR vertex.
The Feynman diagrams pertinent to process (\ref{eq:RRtoHG}) are shown in
Fig.~\ref{fig:RRHg}.

The LO results for the squared amplitudes of subprocesses~(\ref{eq:RRtoH}) and
(\ref{eq:RRtoHG}) that we find by using the Feynman rules of
Ref.~\cite{KTAntonov} coincide with those we obtained in
Ref.~\cite{KniehlSaleevVasin} in the $k_T$-factorization approach.
The general relation between the squared amplitudes in both approaches is
\begin{equation}
{\cal N}\overline{|{\cal A}(R + R \to {\cal H} + X)|^2}
=\overline{|{\cal A}^{\rm KT}(R + R \to {\cal H} + X)|^2}.
\end{equation}
The formulas for the $2\to1$ subprocesses~(\ref{eq:RRtoH}) are listed in
Eq.~(27) of Ref.~\cite{KniehlSaleevVasin}.
In the case of the $2\to2$ subprocess~(\ref{eq:RRtoHG}), our analytic results
were not included in the journal publication of Ref.~\cite{KniehlSaleevVasin}
for lack of space.
However, they are given in Eq.~(38) of the preprint version of
Ref.~\cite{KniehlSaleevVasin} and may be obtained in {\tt FORTRAN} or
{\tt Mathematica} format by electronic mail upon request from the authors.

The differential hadronic cross section of process (\ref{eq:KT}) may then be
evaluated from the squared matrix elements of processes (\ref{eq:RRtoH}) and
(\ref{eq:RRtoHG}) as indicated in Eqs.~(46) and (48) of
Ref.~\cite{KniehlSaleevVasin}.

\section{Bottomonium production at the Tevatron}
\label{sec:five}

The CDF Collaboration at the Tevatron measured the $p_T$ distributions of
$\Upsilon(1S)$, $\Upsilon(2S)$, and $\Upsilon(3S)$ mesons in the central
region of rapidity ($y$), $|y|<0.4$, at $\sqrt{S}=1.8$~TeV (run~I)
\cite{CDFBottomI} and that of the $\Upsilon(1S)$ meson in the rapidity regions
$|y|<0.6$, $0.6<|y|<1.2$, and $1.2<|y|<1.8$ at $\sqrt{S}=1.96$~TeV (run~II)
\cite{CDFBottomII}.
In both cases, the $S$-wave bottomonia were produced promptly, {\it i.e.},
directly or through non-forbidden decays of higher-lying $S$- and $P$-wave
bottomonium states, including cascade transitions such as
$\Upsilon(3S)\to\chi_{b1}(2P)\to\Upsilon(1S)$.

As is well known, the cross section of bottomonium production measured at the
Tevatron at large values of $p_T$ is more than one order of magnitude larger
than the prediction of the color-singlet model (CSM) \cite{CSM} implemented in
the collinear parton model \cite{QWG}.
Switching from the CSM to the NRQCD factorization formalism \cite{NRQCD}
within the collinear parton model \cite{BFL} somewhat ameliorates the
situation in the large-$p_T$ region, at $p_T\agt10$~GeV, but still does not
lead to agreement at all values of $p_T$.
On the other hand, the shape of the $p_T$ distribution can be described in the
color evaporation model \cite{CEM} improved by the resummation of the large
logarithmic contributions from soft-gluon radiation at all orders in
$\alpha_s$ in the region of $p_T<M$ \cite{Berger}.
However, the overall normalization of the cross section can not be predicted
in this approach \cite{CEM,Berger}.

In contrast to previous analyses in the collinear parton model, we perform a
joint fit to the CDF data from run I \cite{CDFBottomI} and run II
\cite{CDFBottomII} for all $p_T$ values, including the small-$p_T$ region, to
extract the color-octet NMEs of the $\Upsilon(nS)$ and $\chi_{bJ}(nP)$ mesons
using three different un-integrated gluon distribution functions.
Our calculations are based on exact analytical expressions for the relevant
squared amplitudes, obtained in the QMRK approach as explained in
Sec.~\ref{sec:four}.

For the reader's convenience, we list in Table~\ref{tab:UpsilonBR}
the inclusive branching fractions of the feed-down decays of the
various bottomonium states, which can be gleaned from
Ref.~\cite{PDG2004}. Theses values supersede those presented in
Ref.~\cite{BFL}. Since the $\Upsilon(nS)$ mesons are identified in
Refs.~\cite{CDFBottomI,CDFBottomII} through their decays to
$\mu^+\mu^-$ pairs, we have to include the corresponding branching
fractions, which we also adopt from Ref.~\cite{PDG2004},
$B(\Upsilon(1S) \to \mu^+ + \mu^-) = 0.0248$, $B(\Upsilon(2S) \to
\mu^+ + \mu^-) = 0.0131$, and $B(\Upsilon(3S) \to \mu^+ + \mu^-) =
0.0181$. We take the pole mass of the bottom quark to be $m_b =
4.77$~GeV.

We now present and discuss our numerical results.
In Table~\ref{tab:NME}, we list our fit results for the relevant color-octet
NMEs for three different choices of un-integrated gluon distribution function,
namely JB \cite{JB}, JS \cite{JS}, and KMR \cite{KMR}.
The relevant color-singlet NMEs are not fitted.
The color-singlet NMEs of the $\Upsilon(nS)$ mesons are determined from the
measured partial decay widths of $\Upsilon(nS) \to l^+ + l^-$ using the vacuum
saturation approximation and heavy-quark spin symmetry in the NRQCD
factorization formulas and including NLO QCD radiative corrections
\cite{QCDCorrections}.
The partial decay widths of $\chi_{b0}(nP)\to 2\gamma$, from which the
color-singlet NMEs of the $\chi_{bJ}(nP)$ mesons could be extracted, are yet
unknown.
However, these NMEs can be estimated using the wave functions evaluated at the
origin from potential models \cite{QPM}, as was done in Ref.~\cite{BFL}.
We adopt the color-singlet NMEs of the $\chi_{b0}(nP)$ mesons from
Ref.~\cite{BFL}.

\begin{figure}[hpt]
\begin{center}
\psfig{figure=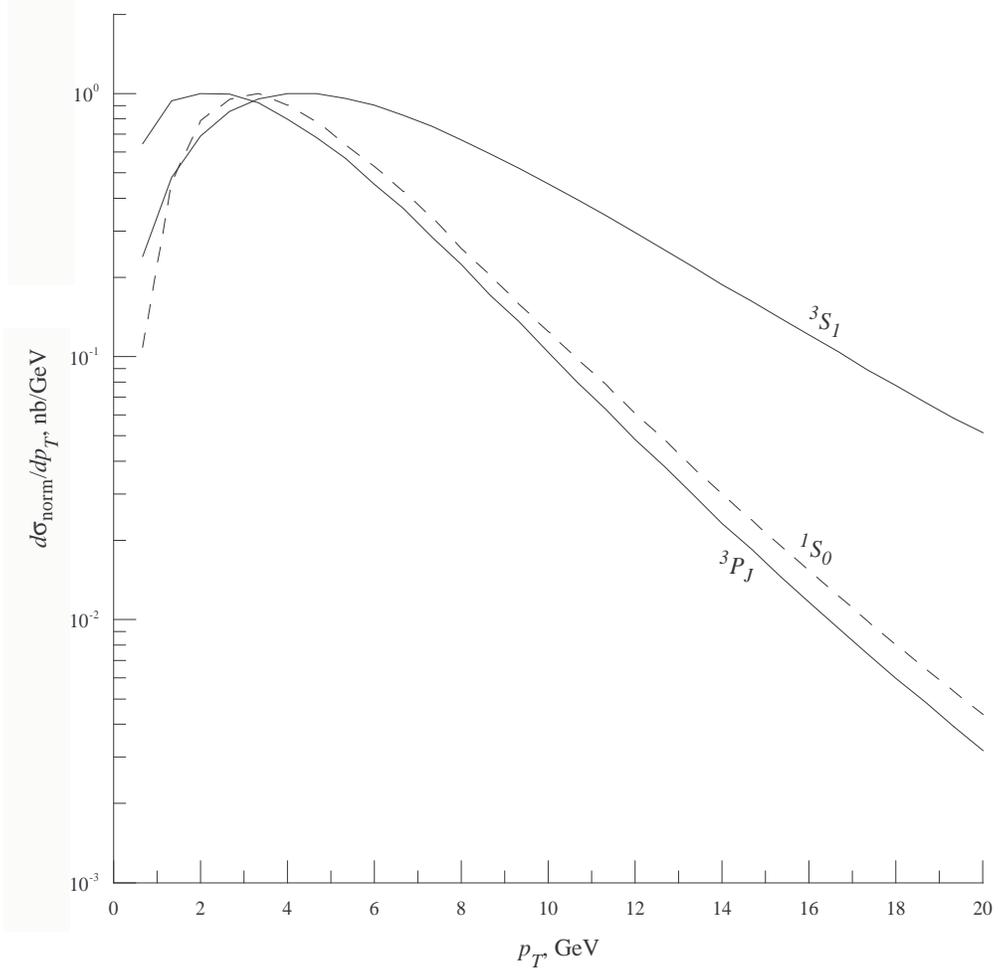,width=0.8\textwidth}
\caption{\label{fig:States}Contributions to the $p_T$ distribution of direct
$\Upsilon(1S)$ hadroproduction in $p\bar p$ scattering with $\sqrt{S}=1.8$~TeV
and $|y|<0.4$ from the relevant color-octet states.
All distributions are normalized to unity at their peaks.}
\end{center}
\end{figure}

We first study the relative importance of the various color-octet $b\bar b$
Fock states in direct $\Upsilon(nS)$ hadroproduction.
Previous fits to CDF data \cite{CDFBottomI} were constrained to the
large-$p_T$ region, $p_T\agt8$~GeV, and could not separate the contributions
proportional to $\langle {\cal O}^{\Upsilon(nS)}[^1S_0^{(8)}]\rangle$
and $\langle {\cal O}^{\Upsilon(nS)}[^3P_0^{(8)}]\rangle$.
Instead, they determined the linear combination
\begin{equation}
M_r^{\Upsilon(nS)}
=\langle{\cal O}^{\Upsilon(nS)}\left[^1S_0^{(8)}\right]\rangle
+\frac{r}{m_b^2}\langle{\cal O}^{\Upsilon(nS)}\left[^3P_0^{(8)}\right]\rangle,
\label{eq:lc}
\end{equation}
for the value of $r$ that minimized the error on $M_r^{\Upsilon(nS)}$.
By contrast, the QMRK approach allows us to cover also the small-$p_T$ region
and thus to fit $\langle {\cal O}^{\Upsilon(nS)}[^1S_0^{(8)}]\rangle$ and
$\langle {\cal O}^{\Upsilon(nS)}[^3P_0^{(8)}]\rangle$ separately, thanks to
the different $p_T$ dependences of the respective contributions for
$p_T\alt8$~GeV.
This feature is nicely illustrated for direct $\Upsilon(1S)$ hadroproduction
in Fig.~\ref{fig:States}, where the shapes of the contributions proportional
to $\langle {\cal O}^{\Upsilon(1S)}[^3S_1^{(8)}]\rangle$,
$\langle {\cal O}^{\Upsilon(1S)}[^1S_0^{(8)}]\rangle$, and
$\langle{\cal O}^{\Upsilon(1S)}[^3P_0^{(8)}]\rangle$ are compared.
Notice that the peak positions significantly differ, by up to 2~GeV.
Apparently, this suffices to disentangle the contributions previously
combined by Eq.~(\ref{eq:lc}).

\begin{figure}[hpt]
\begin{center}
\psfig{figure=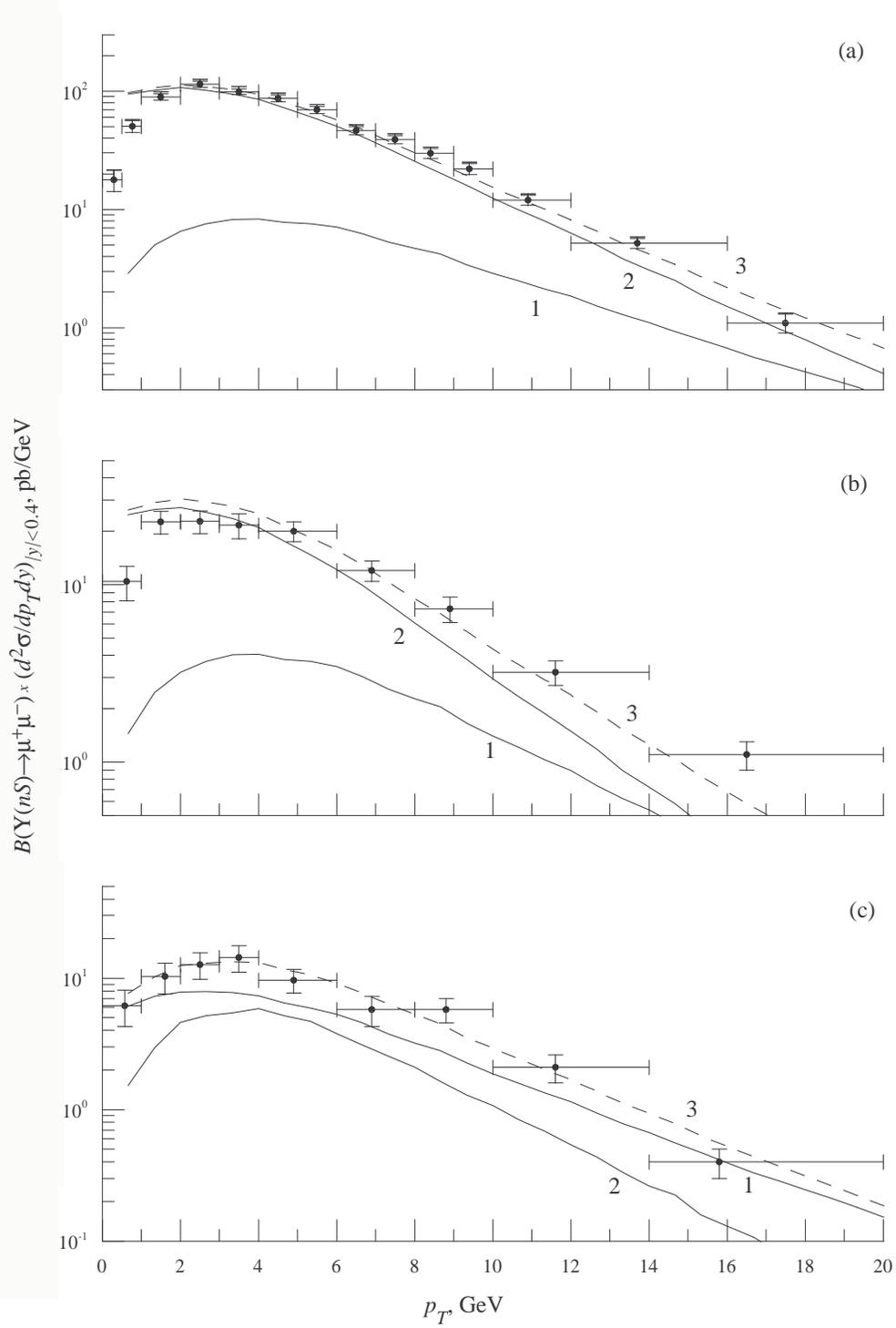,width=0.8\textwidth}
\caption{\label{fig:UpsilonR12JB}$p_T$ distributions of prompt (a)
$\Upsilon(1S)$, (b) $\Upsilon(2S)$, and (c) $\Upsilon(3S)$ hadroproduction in
$p\overline{p}$ scattering with $\sqrt{S}=1.8$~TeV and $|y|<0.4$ including the
respective decay branching fractions $B(\Upsilon(nS)\to\mu^++\mu^-)$.
The color-octet (curve 1) and color-singlet (curve 2) contributions, evaluated
with the JB \cite{JB} un-integrated gluon distribution function, and their
sum (curve 3) are compared with the CDF data from run I \cite{CDFBottomI}.}
\end{center}
\end{figure}

\begin{figure}[hpt]
\begin{center}
\psfig{figure=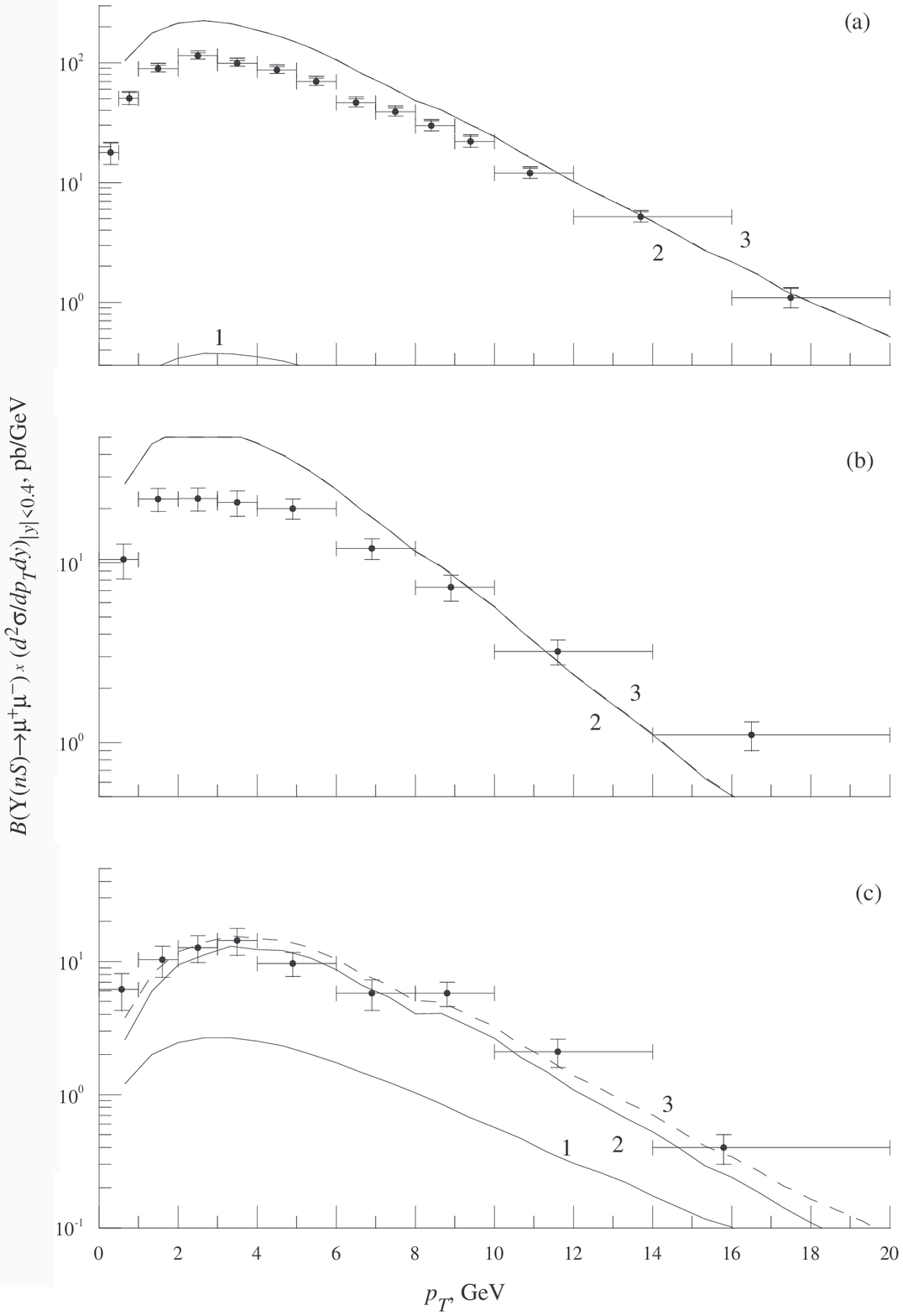,width=0.8\textwidth}
\caption{\label{fig:UpsilonR12JS}Same as in Fig.~\ref{fig:UpsilonR12JB}, but
for the JS \cite{JS} un-integrated gluon distribution function.}
\end{center}
\end{figure}

\begin{figure}[hpt]
\begin{center}
\psfig{figure=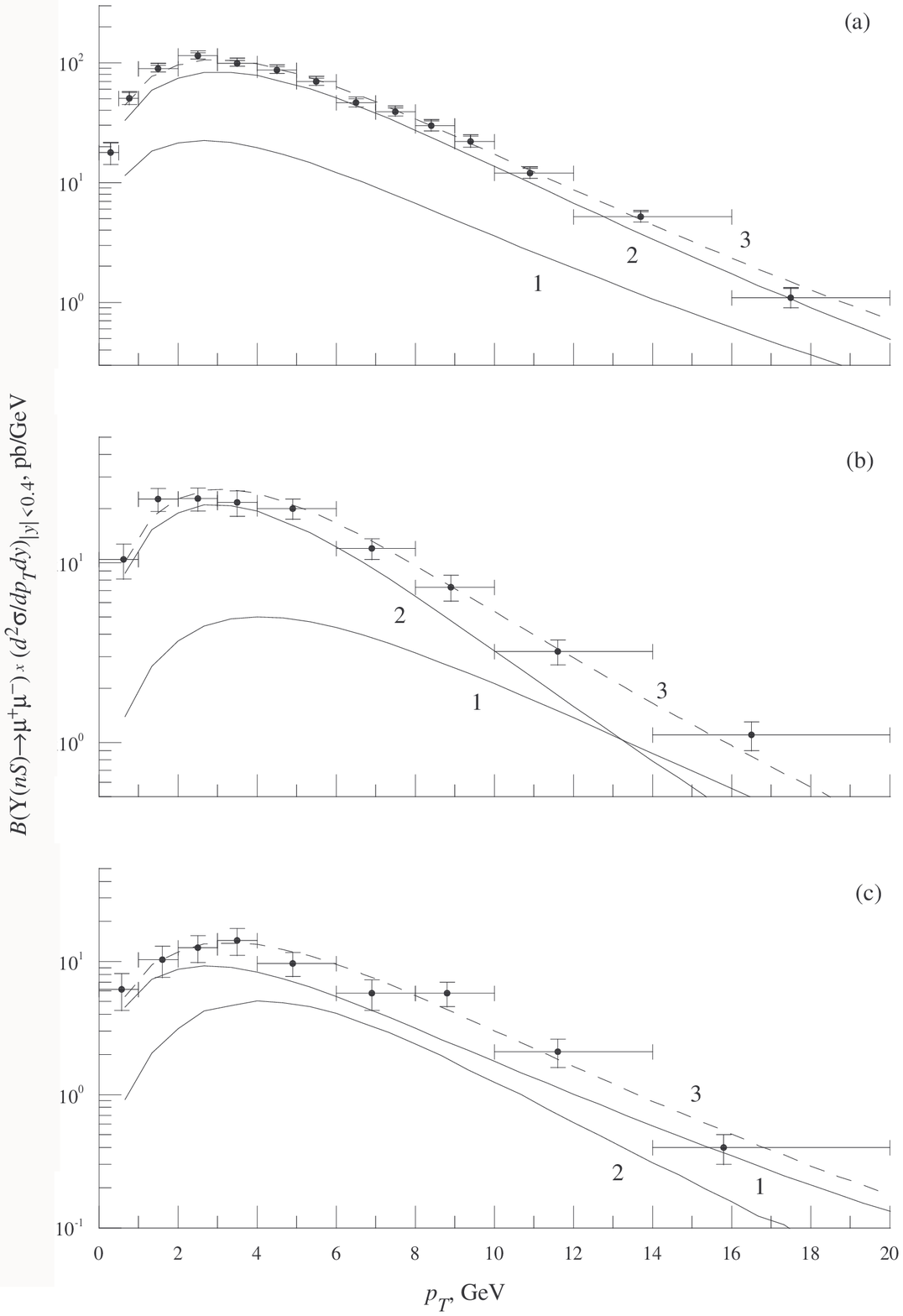,width=0.8\textwidth}
\caption{\label{fig:UpsilonR12KMR}Same as in Fig.~\ref{fig:UpsilonR12JB}, but
for the KMR \cite{KMR} un-integrated gluon distribution function.}
\end{center}
\end{figure}

In Figs.~\ref{fig:UpsilonR12JB}, \ref{fig:UpsilonR12JS}, and
\ref{fig:UpsilonR12KMR}, we compare the CDF data on prompt
$\Upsilon(nS)$ hadroproduction in run I \cite{CDFBottomI} with the theoretical
results evaluated with the JB \cite{JB}, JS \cite{JS}, and KMR \cite{KMR}
un-integrated gluon distribution functions, respectively, and the NMEs listed
in Table~\ref{tab:NME}.
In each case, the color-singlet and color-octet contributions are also shown
separately.
Except for the JB and KMR analyses of $\Upsilon(3S)$ production, the
color-octet contributions are always suppressed, especially at low values of
$p_T$.
In the JS analysis, the $\Upsilon(1S)$ and $\Upsilon(2S)$ data are
significantly exceeded by the color-singlet contributions for $p_T\alt10$~GeV,
which explains the poor quality of the fit, with $\chi^2/\mbox{d.o.f.}=27$.
In the JB analysis, this only happens for $p_T\alt2$~GeV, so that the value of
$\chi^2/\mbox{d.o.f.}$ is lowered by one order of magnitude, being
$\chi^2/\mbox{d.o.f.}=2.9$.
By contrast, the KMR gluon yields an excellent fit, with
$\chi^2/\mbox{d.o.f.}=0.5$, and will be the only one considered in the
following discussion.
Comparing the color-singlet and color-octet contributions in
Fig.~\ref{fig:UpsilonR12KMR}, we observe that the latter is dominant in the
$\Upsilon(3S)$ case and in the $\Upsilon(2S)$ case for $p_T\agt13$~GeV, while
it is of minor importance in the $\Upsilon(1S)$ case in the whole $p_T$ range
considered.
The latter feature is substantiated by the run-II data and is reflected in all
their $y$ subintervals, as may be see from Fig.~\ref{fig:CDFBottomII}.

\begin{figure}[hpt]
\begin{center}
\psfig{figure=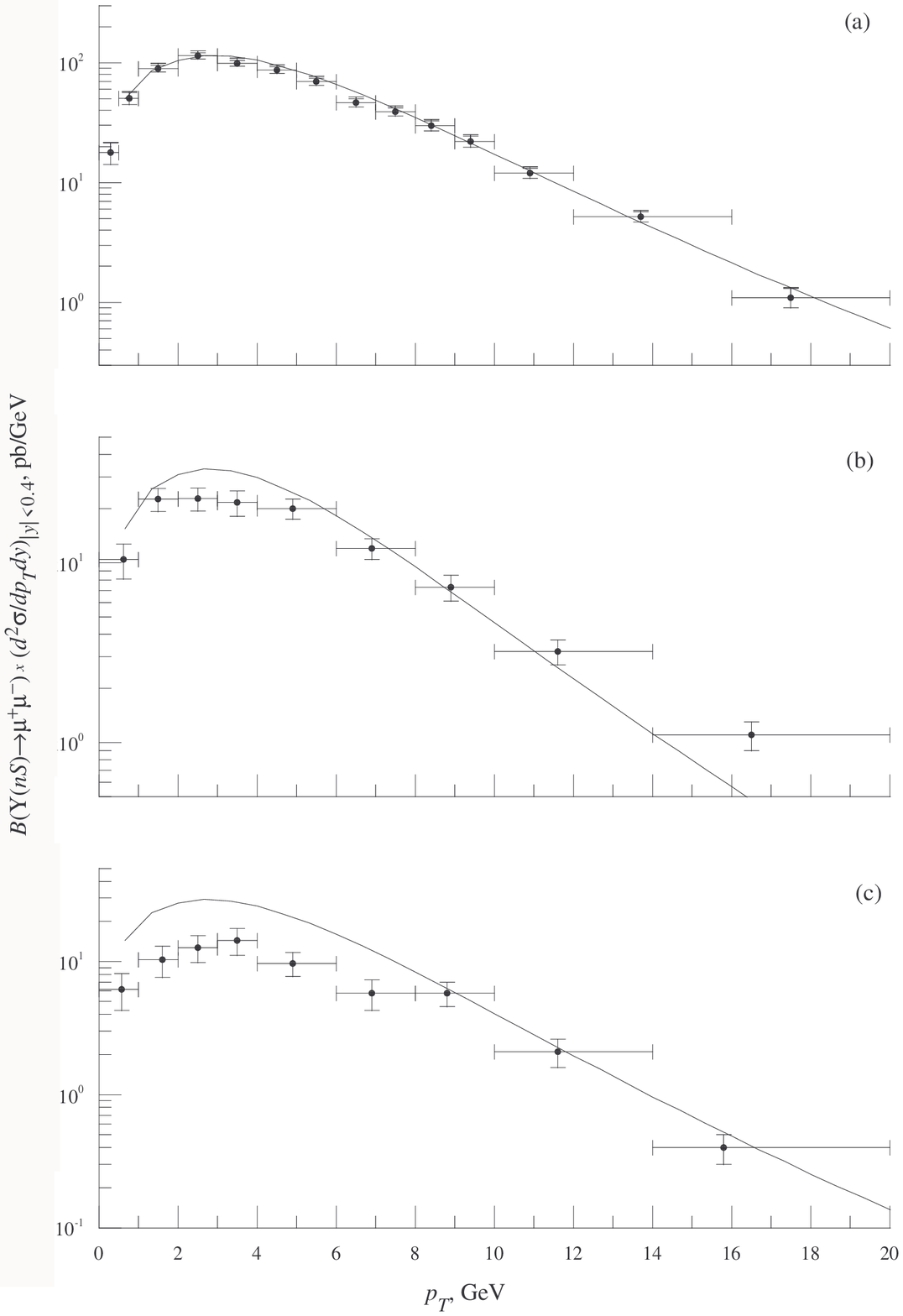,width=0.8\textwidth}
\caption{\label{fig:UpsilonR12CSM}$p_T$ distributions of prompt (a)
$\Upsilon(1S)$, (b) $\Upsilon(2S)$, and (c) $\Upsilon(3S)$ hadroproduction in
$p\overline{p}$ scattering with $\sqrt{S}=1.8$~TeV and $|y|<0.4$ including the
respective decay branching fractions $B(\Upsilon(nS)\to\mu^++\mu^-)$.
The color-singlet contribution including the estimated feed-down contributions
due to the $\chi_{bJ}(3P)$ meson, evaluated with the KMR \cite{KMR}
un-integrated gluon distribution function, is compared with the CDF data from
run I \cite{CDFBottomI}.}
\end{center}
\end{figure}

Notice that the contributions to prompt $\Upsilon(nS)$ hadroproduction due to
the feed-down from the $\chi_{bJ}(3P)$ mesons have been neglected above,
simply because the latter have not yet been observed and their partial decay
widths are unknown.
In the remainder of this section, we assess the impact of these contributions.
For the color-singlet NME, we use the potential model result
$\langle {\cal O}^{\chi_{b0}(3P)}[^3P_0^{(1)}]\rangle=2.7$~GeV$^5$ \cite{QPM}.
By analogy to the KMR fit results for
$\langle {\cal O}^{\chi_{b0}(1P)}[^3S_1^{(8)}]\rangle$ and
$\langle {\cal O}^{\chi_{b0}(2P)}[^3S_1^{(8)}]\rangle$ in Table~\ref{tab:NME},
we expect the value of $\langle {\cal O}^{\chi_{b0}(3P)}[^3S_1^{(8)}]\rangle$
to be negligibly small, compatible with zero.
Looking at Table~\ref{tab:UpsilonBR}, a naive extrapolation from the
$\chi_{bJ}(1P)$ and $\chi_{bJ}(2P)$ states suggests that the inclusive
branching fractions for the $\chi_{bJ}(3P)$ decays into the $\Upsilon(3S)$,
$\Upsilon(2S)$, and $\Upsilon(1S)$ states could be about 12\%, 9\%, and 7\%,
respectively.
These decays generate further cascade transitions, whose inclusive
branching fractions follow from these estimates in combination with the
entries of Table~\ref{tab:UpsilonBR}.
Including all these ingredients, we repeat our KMR fit to the CDF data.
As illustrated in Fig.~\ref{fig:UpsilonR12CSM} for prompt $\Upsilon(nS)$
hadroproduction in run~I, the CDF data can be fairly well described in the
QMRK approach to the CSM, while the color-octet contributions turn out to be
negligibly small.
We note in passing that a similar observation, although with lower degree of
agreement between data and theory, can be made for the JB gluon, while the JS
gluon badly fails for $p_T\alt10$~GeV.

\section{Conclusion}
\label{sec:six}

Working at LO in the QMRK approach to NRQCD, we analytically evaluated the
squared amplitudes of prompt bottomonium production in two-reggeon collisions.
We extracted the relevant color-octet NMEs,
$\langle {\cal O}^{\cal H}[^3S_1^{(8)}]\rangle$,
$\langle {\cal O}^{\cal H}[^1S_0^{(8)}]\rangle$, and
$\langle {\cal O}^{\cal H}[^3P_0^{(8)}]\rangle$ for
${\cal H}=\Upsilon(1S)$, $\Upsilon(2S)$, $\Upsilon(3S)$, $\chi_{b0}(1P)$, and
$\chi_{b0}(2P)$, through fits to $p_T$ distributions of prompt $\Upsilon(nS)$
hadroproduction measured by the CDF Collaboration at the Tevatron in $p\bar p$
collisions with $\sqrt{S}=1.8$~TeV \cite{CDFBottomI} and 1.96~TeV
\cite{CDFBottomII} using three different un-integrated gluon distribution
functions of the proton, namely JB \cite{JB}, JS \cite{JS}, and KMR
\cite{KMR}.
The fits based on the KMR, JB, and JS gluons turned out to be excellent, fair,
and poor, respectively.
They yielded small to vanishing values for the color-octet NMEs, especially
when the estimated feed-down contributions from the as-yet unobserved
$\chi_{bJ}(3P)$ states were included.

The present analysis, together with a recent investigation of charmonium
production at high energies \cite{KniehlSaleevVasin}, suggest that the
color-octet NMEs of bottomonium are more strongly suppressed than those of
charmonium as expected from the velocity scaling rules of NRQCD.
We illustrated that the QMRK approach \cite{KTLipatov,KTAntonov} provides a
useful laboratory to describe the phenomenology of high-energy processes in
the Regge limit of QCD.

LO predictions in both the collinear parton model and the QMRK framework
suffer from sizeable theoretical uncertainties, which are largely due to
unphysical-scale dependences.
Substantial improvement can only be achieved by performing full NLO analyses.
While the stage for the NLO NRQCD treatment of $2\to2$ processes has been set
in the collinear parton model \cite{kkms}, conceptual issues still remain to
be elaborated in the QMRK approach.
Since, at NLO, incoming partons can gain a finite $k_T$ kick through the
perturbative emission of partons, one expects that essential features produced
by the QMRK approach at LO will thus automatically show up at NLO in the
collinear parton model.

\section{Acknowledgements}

We thank E.~Kuraev and M.~Ryskin for useful discussions.
D.V.V. is grateful to the International Center of Fundamental Physics in
Moscow and the Dynastiya Foundation for financial support.
This work was supported in part by BMBF Grant No.\ 05 HT4GUA/4 and by DFG
Grant No.\ KN~365/6--1.

\newpage

\begin{table}[hpt]
\begin{center}
\caption{\label{tab:UpsilonBR}Inclusive branching fractions of the feed-down
decays of the various bottomonium states.}
\begin{ruledtabular}
\begin{tabular}{c|ccccccccc}
In$\setminus$Out & $\Upsilon(3S)$ & $\chi_{b2}(2P)$ & $\chi_{b1}(2P)$ & 
$\chi_{b0}(2P)$ & $\Upsilon(2S)$ & $\chi_{b2}(1P)$ & $\chi_{b1}(1P)$ & 
$\chi_{b0}(1P)$ & $\Upsilon(1S)$ \\
\hline
$\Upsilon(3S)$ & 1 & 0.114 & 0.113 & 0.054 & 0.106 & 0.00721 & 0.00742 & 
0.00403 & 0.102 \\
$\chi_{b2}(2P)$ & $\cdots$ & 1 & $\cdots$ & $\cdots$ & 0.162 & 0.0110 & 
0.0113 & 0.00616 & 0.130 \\
$\chi_{b1}(2P)$ & $\cdots$ & $\cdots$ & 1 & $\cdots$ & 0.21 & 0.0143 & 
0.0147 & 0.00798 & 0.161 \\
$\chi_{b0}(2P)$ & $\cdots$ & $\cdots$ & $\cdots$ & 1 & 0.046 & 0.00313 & 
0.00322 & 0.00175 & 0.0167 \\
$\Upsilon(2S)$ & $\cdots$ & $\cdots$ & $\cdots$ & $\cdots$ & 1 & 0.068 & 
0.07 & 0.038 & 0.320 \\
$\chi_{b2}(1P)$ & $\cdots$ & $\cdots$ & $\cdots$ & $\cdots$ & $\cdots$ & 
1 & $\cdots$ & $\cdots$ & 0.22 \\
$\chi_{b1}(1P)$ & $\cdots$ & $\cdots$ & $\cdots$ & $\cdots$ & $\cdots$ & 
$\cdots$ & 1 & $\cdots$ & 0.35 \\
$\chi_{b0}(1P)$ & $\cdots$ & $\cdots$ & $\cdots$ & $\cdots$ & $\cdots$ & 
$\cdots$ & $\cdots$ & 1 & 0.06 \\
$\Upsilon(1S)$ & $\cdots$ & $\cdots$ & $\cdots$ & $\cdots$ & $\cdots$ & 
$\cdots$ & $\cdots$ & $\cdots$ & 1 \\
\end{tabular}
\end{ruledtabular}
\end{center}
\end{table}

\newpage

\begin{table}[hpt]
\begin{center}
\caption{\label{tab:NME}NMEs of the $\Upsilon(1S)$, $\Upsilon(2S)$,
$\Upsilon(3S)$, $\chi_{b0}(1P)$, and $\chi_{b0}(2P)$ mesons from fits to CDF
data from run I \cite{CDFBottomI} and run II \cite{CDFBottomII} in the
collinear parton model (PM) \cite{BFL} using the CTEQ5L \cite{CTEQ} parton
distribution functions of the proton and in the QMRK approach using the JB
\cite{JB}, JS \cite{JS}, and KMR \cite{KMR} un-integrated gluon distribution
functions of the proton.
The errors on the fit results are determined by varying in turn each NME up
and down about its central value until the value of $\chi^2$ is increased by
unity keeping all other NMEs fixed at their central values.}
\begin{ruledtabular}
\begin{tabular}{ccccc}
NME & PM \cite{BFL} & Fit JB & Fit JS & Fit KMR \\
\hline
$\langle{\cal O}^{\Upsilon(1S)}[^3S_1^{(1)}]\rangle/$GeV$^3$ &
$10.9\pm 1.6$ & $10.9\pm 1.6$ &  $10.9\pm 1.6$ & $10.9\pm 1.6$ \\
$\langle{\cal O}^{\Upsilon(1S)}[^3S_1^{(8)}]\rangle/$GeV$^3$ &
$\left(2.0\pm 4.1{-0.6\choosen{+0.5}}\right)\times 10^{-2}$ &
$(5.3\pm 0.5)\times 10^{-3}$ & $(0.0\pm 1.8)\times 10^{-4}$ &
$(0.0\pm 3.1)\times 10^{-3}$ \\
$\langle{\cal O}^{\Upsilon(1S)}[^1S_0^{(8)}]\rangle/$GeV$^3$ &
$\cdots$ & $(0.0\pm 4.7)\times 10^{-4}$ & $(0.0\pm 5.2)\times
10^{-5}$ &
$(0.0\pm 4.3)\times 10^{-3}$ \\
$\langle{\cal O}^{\Upsilon(1S)}[^3P_0^{(8)}]\rangle/$GeV$^5$ &
$\cdots$ & $(0.0\pm 1.3)\times 10^{-3}$ & $(0.0\pm 1.6)\times
10^{-4}$ &
$(9.5\pm 2.0)\times 10^{-2}$ \\
$M_{5}^{\Upsilon(1S)}/$GeV$^3$ &
$\left(1.4\pm 0.7{+1.0\choosen{-0.7}}\right)\times 10^{-1}$ &
$(0.0\pm 7.6)\times 10^{-4}$ & $(0.0\pm 8.7)\times 10^{-5}$ &
$(2.1\pm 0.9)\times 10^{-2}$\\
%\hline
$\langle{\cal O}^{\chi_{b0}(1P)}[^3P_0^{(1)}]\rangle/$GeV$^5$ &
$2.4\pm 0.4$ & $2.4\pm 0.4$ & $2.4\pm 0.4$ & $2.4\pm 0.4$ \\
$\langle{\cal O}^{\chi_{b0}(1P)}[^3S_1^{(8)}]\rangle/$GeV$^3$ &
$\left(1.5\pm 1.1{+1.3\choosen{-1.0}}\right)\times 10^{-2}$ &
$(0.0\pm 2.1)\times 10^{-3}$ & $(0.0\pm 8.4)\times 10^{-5}$ &
$(0.0\pm 1.4)\times 10^{-3}$ \\
%\hline
$\langle{\cal O}^{\Upsilon(2S)}[^3S_1^{(1)}]\rangle/$GeV$^3$ &
$4.5\pm 0.7$ & $4.5\pm 0.7$ & $4.5\pm 0.7$ & $4.5\pm 0.7$ \\
$\langle{\cal O}^{\Upsilon(2S)}[^3S_1^{(8)}]\rangle/$GeV$^3$ &
$\left(1.6\pm 0.6{+0.7\choosen{-0.5}}\right)\times 10^{-1}$ &
$(0.0\pm 5.9)\times 10^{-3}$ & $(0.0\pm 4.1)\times 10^{-4}$ &
$(3.3\pm 0.8)\times 10^{-2}$ \\
$\langle{\cal O}^{\Upsilon(2S)}[^1S_0^{(8)}]\rangle/$GeV$^3$ &
$\cdots$ & $(0.0\pm 9.2)\times 10^{-4}$ & $(0.0\pm 8.3)\times
10^{-5}$ &
$(0.0\pm 3.7)\times 10^{-3}$ \\
$\langle{\cal O}^{\Upsilon(2S)}[^3P_0^{(8)}]\rangle/$GeV$^5$ &
$\cdots$ & $(0.0\pm 2.6)\times 10^{-3}$ & $(0.0\pm 2.8)\times
10^{-4}$ &
$(0.0\pm 1.6)\times 10^{-2}$ \\
$M_{5}^{\Upsilon(2S)}/$GeV$^3$ &
$\left(-1.1\pm 1.0 {+0.3\choosen{-0.2}}\right)\times 10^{-1}$ &
$(0.0\pm 1.5)\times 10^{-3}$ & $(0.0\pm 1.4)\times 10^{-4}$ &
$(0.0\pm 7.2)\times 10^{-3}$ \\
%\hline
$\langle{\cal O}^{\chi_{b0}(2P)}[^3P_0^{(1)}]\rangle/$GeV$^5$ &
$2.6\pm 0.5$ & $2.6\pm 0.5$ & $2.6\pm 0.5$ & $2.6\pm 0.5$ \\
$\langle{\cal O}^{\chi_{b0}(2P)}[^3S_1^{(8)}]\rangle/$GeV$^3$ &
$\left(0.8\pm 1.1{+1.1\choosen{-0.8}}\right)\times 10^{-2}$ &
$(1.1\pm 0.4)\times 10^{-2}$ & $(0.0\pm 2.8)\times 10^{-4}$ &
$(0.0\pm 5.7)\times 10^{-3}$ \\
%\hline
$\langle{\cal O}^{\Upsilon(3S)}[^3S_1^{(1)}]\rangle/$GeV$^3$ &
$4.3\pm 0.9$ & $4.3\pm 0.9$ & $4.3\pm 0.9$ & $4.3\pm 0.9$ \\
$\langle{\cal O}^{\Upsilon(3S)}[^3S_1^{(8)}]\rangle/$GeV$^3$ &
$\left(3.6\pm 1.9 {+1.8\choosen{-1.3}}\right)\times 10^{-2}$ &
$(1.4\pm 0.3)\times 10^{-2}$ & $(5.9\pm 4.2)\times 10^{-3}$ &
$(1.1\pm 0.4)\times 10^{-2}$ \\
$\langle{\cal O}^{\Upsilon(3S)}[^1S_0^{(8)}]\rangle/$GeV$^3$ &
$\cdots$ & $(0.0\pm 2.6)\times 10^{-3}$ & $(0.0\pm 8.1)\times
10^{-4}$ &
$(0.0\pm 2.7)\times 10^{-3}$ \\
$\langle{\cal O}^{\Upsilon(3S)}[^3P_0^{(8)}]\rangle/$GeV$^5$ &
$\cdots$ & $(2.4\pm 0.8)\times 10^{-2}$ & $(3.4\pm 4.2)\times
10^{-3}$ &
$(5.2\pm 1.1)\times 10^{-2}$ \\
$M_{5}^{\Upsilon(3S)}/$GeV$^3$ &
$\left(5.4\pm 4.3 {+3.1\choosen{-2.2}}\right)\times 10^{-2}$ &
$(5.2\pm 4.4)\times 10^{-3}$ & $(7.4\pm 10.2)\times 10^{-4}$ &
$(1.1\pm 0.5)\times 10^{-2}$\\
%\hline
$\langle{\cal O}^{\chi_{b0}(3P)}[^3P_0^{(1)}]\rangle/$GeV$^5$ &
$2.7\pm 0.7$ & $2.7\pm 0.7$ & $2.7\pm 0.7$ & $2.7\pm 0.7$ \\
%\hline
$\chi^2/\mathrm{d.o.f.}$ & $\cdots$  & $2.9$ & $27$ & $0.5$ \\
\end{tabular}
\end{ruledtabular}
\end{center}
\end{table}

\newpage

\begin{figure}[hpt]
\begin{center}
\psfig{figure=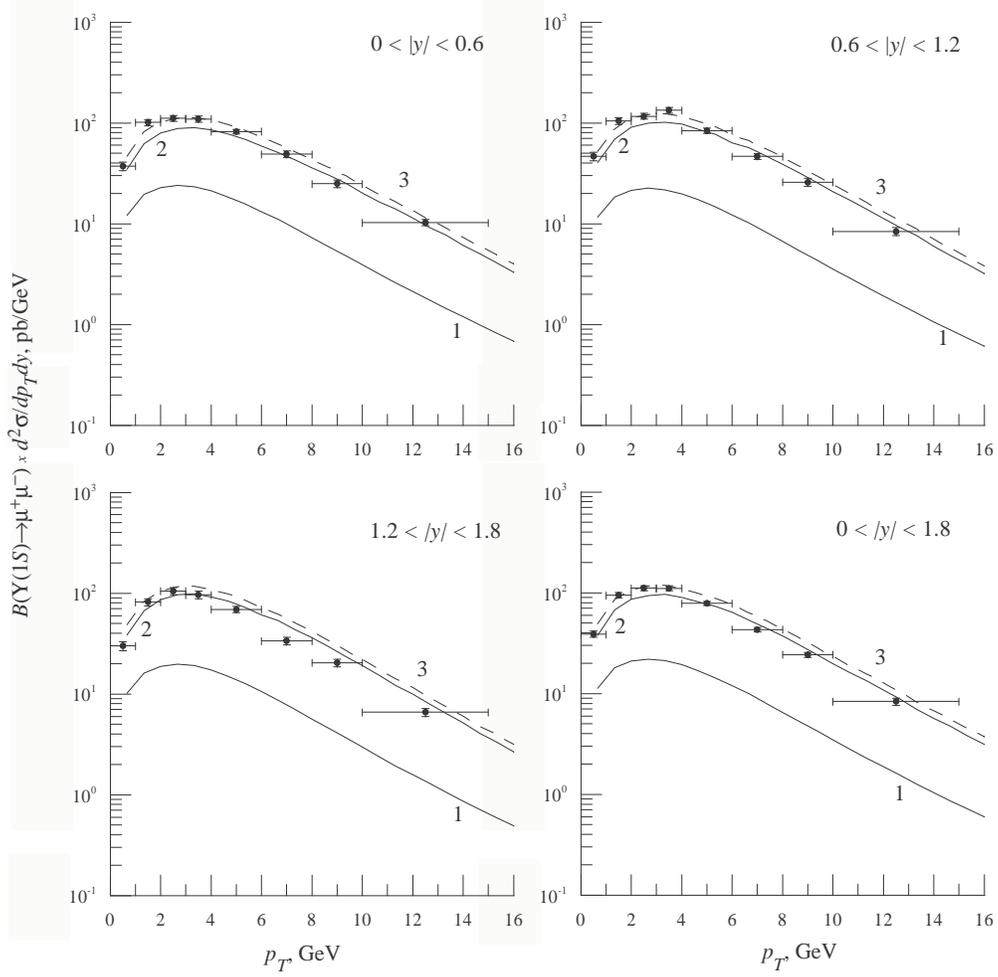,width=0.8\textwidth}
\caption{\label{fig:UpsilonR21}$p_T$ distributions of prompt $\Upsilon(1S)$
hadroproduction in $p\overline{p}$ scattering with $\sqrt{S}=1.96$~TeV and (a)
$|y|<0.6$, (b) $0.6<|y|<1.2$, (c) $1.2<|y|<1.8$, and (d) $|y|<1.8$ including
the decay branching fractions $B(\Upsilon(1S)\to\mu^++\mu^-)$.
The color-octet (curve 1) and color-singlet (curve 2) contributions, evaluated
with the KMR \cite{KMR} un-integrated gluon distribution function, and their
sum (curve 3) are compared with the CDF data from run II \cite{CDFBottomII}.}
\end{center}
\end{figure}

\begin{thebibliography}{99}

\bibitem{NRQCD}
G.~T.~Bodwin, E.~Braaten, and G.~P.~Lepage,
Phys.\ Rev.\ D \textbf{51}, 1125 (1995); \textbf{55}, 5853(E) (1997).

\bibitem{PartonModel}
CTEQ Collaboration, R. Brock \emph{et al.},
Rev.\ Mod.\ Phys.\ \textbf{67}, 157 (1995).

\bibitem{DGLAP}
V.~N.~Gribov and L.~N.~Lipatov,
Sov.\ J.\ Nucl.\ Phys.\ \textbf{15}, 438 (1972)
[Yad.\ Fiz.\ \textbf{15}, 781 (1972)];
Yu.~L.~Dokshitzer,
Sov.\ Phys.\ JETP \textbf{46}, 641 (1977)
[Zh.\ Eksp.\ Teor.\ Fiz.\  \textbf{73}, 1216 (1977)];
G.~Altarelli and G.~Parisi,
Nucl.\ Phys.\ \textbf{B126}, 298 (1977).

\bibitem{BFKL}
E.~A.~Kuraev, L.~N.~Lipatov, and V.~S.~Fadin,
Sov.\ Phys.\ JETP \textbf{44}, 443 (1976)
[Zh.\ Eksp.\ Teor.\ Fiz.\  \textbf{71}, 840 (1976)];
I.~I.~Balitsky and L.~N.~Lipatov,
Sov.\ J.\ Nucl.\ Phys.\  \textbf{28}, 822 (1978)
[Yad.\ Fiz.\ \textbf{28}, 1597 (1978)].

\bibitem{KTGribov}
L.~V.~Gribov, E.~M.~Levin, and M.~G.~Ryskin,
Phys.\ Rept.\ \textbf{100}, 1 (1983);
S.~Catani, M.~Ciafoloni, and F.~Hautmann,
Nucl.\ Phys.\ \textbf{B366}, 135 (1991).

\bibitem{KTCollins}
J.~C.~Collins and R.~K.~Ellis,
Nucl.\ Phys.\ \textbf{B360}, 3 (1991).

\bibitem{KTLipatovFadin}
V.~S.~Fadin and L.~N.~Lipatov,
Nucl.\ Phys.\ \textbf{B477}, 767 (1996).

\bibitem{KTAntonov}
E.~N.~Antonov, L.~N.~Lipatov, E.~A.~Kuraev, and I.~O.~Cherednikov,
Nucl.\ Phys.\ \textbf{B721}, 111 (2005).

\bibitem{KTLipatov}
L.~N.~Lipatov,
Nucl.\ Phys.\ \textbf{B452}, 369 (1995).

\bibitem{PRD2003}
V.~A.~Saleev and D.~V.~Vasin,
Phys.\ Rev.\ D \textbf{68}, 114013 (2003);
Phys.\ Atom.\ Nucl.\ \textbf{68}, 94 (2005)
[Yad.\ Fiz.\ \textbf{68}, 95 (2005)].

\bibitem{KniehlSaleevVasin}
B.~A.~Kniehl, D.~V.~Vasin, and V.~A.~Saleev,
Phys.\ Rev.\ D \textbf{73}, 074022 (2006).

\bibitem{smallx}
Small $x$ Collaboration, B.~Anderson \emph{et al.},
Eur.\ Phys.\ J. C \textbf{25}, 77 (2002).

\bibitem{Ostrovsky}
V.~S.~Fadin, M.~I.~Kotsky, and L.~N.~Lipatov,
Phys.\ Lett.\ B \textbf{415}, 97 (1997);
A.~Leonidov and D.~Ostrovsky,
Eur.\ Phys.\ J. C\ \textbf{11}, 495 (1999);
D.~Ostrovsky,
Phys.\ Rev.\ D \textbf{62}, 054028 (2000);
V.~S.~Fadin, M.~G.~Kozlov, and A.~V.~Reznichenko,
Phys.\ Atom.\ Nucl.\ \textbf{67}, 359 (2004)
[Yad.\ Fiz.\ \textbf{67}, 377 (2004)].

\bibitem{JB}
J.~Bl\"umlein,
Report No.\ DESY~95--121 (1995).

\bibitem{JS}
H.~Jung and G.~P.~Salam,
Eur.\ Phys.\ J. C \textbf{19}, 351 (2001).

\bibitem{KMR}
M.~A.~Kimber, A.~D.~Martin, and M.~G.~Ryskin,
Phys.\ Rev.\ D \textbf{63}, 114027 (2001).

\bibitem{PLB2002}
V.~A.~Saleev and D.~V.~Vasin,
Phys.\ Lett.\ B \textbf{548}, 161 (2002).

\bibitem{CCFM}
M.~Ciafaloni,
Nucl.\ Phys.\ \textbf{B296}, 49 (1988);
S.~Catani, F.~Fiorani, and G.~Marchesini,
Phys.\ Lett.\ B \textbf{234}, 339 (1990);
G.~Marchesini,
Nucl.\ Phys.\ \textbf{B445}, 49 (1995).

\bibitem{CDFBottomI}
CDF Collaboration, F.~Abe \emph{et al.},
Phys.\ Rev.\ Lett.\ \textbf{75}, 4358 (1995);
CDF Collaboration, D.~Acosta \emph{et al.},
\emph{ibid.}\ \textbf{88}, 161802 (2002).

\bibitem{CDFBottomII}
CDF Collaboration, V.~M.~Abazov \emph{et al.},
Phys.\ Rev.\ Lett.\ \textbf{94}, 232001 (2005).

\bibitem{CSM}
V.~G.~Kartvelishvili, A.~K.~Likhoded, and S.~R.~Slabospitsky,
Sov.\ J. Nucl.\ Phys.\ \textbf{28}, 678 (1978)
[Yad.\ Fiz.\ \textbf{28}, 1315 (1978)];
E.~L.~Berger and D.~Jones,
Phys.\ Rev.\ D \textbf{23}, 1521 (1981);
R.~Baier and R.~R\"{u}ckl,
Phys.\ Lett.\ B \textbf{102}, 364 (1981).

\bibitem{QWG}
N.~Brambilla \emph{et al.},
CERN Yellow Report No.\ CERN-2005-005 and No.\ FERMILAB-FN-0779, 2005.

\bibitem{BFL}
E.~Braaten, S.~Fleming, and A.~K.~Leibovich,
Phys.\ Rev.\ D \textbf{63}, 094006 (2001).

\bibitem{CEM}
J.~F.~Amundson, O.~J.~P.~Eboli, E.~M.~Gregores, and F.~Halzen,
Phys.\ Lett.\ B \textbf{390}, 323 (1997).

\bibitem{Berger}
E.~L.~Berger, J.~Qiu, and Y.~Wang,
Phys.\ Rev.\ D \textbf{71}, 034007 (2005);
Int.\ J. Mod.\ Phys.\ A \textbf{20}, 3753 (2005).

\bibitem{PDG2004}
Particle Data Group, S.~Eidelman \emph{et al.},
Phys.\ Lett.\ B \textbf{592}, 1 (2004).

\bibitem{CTEQ}
CTEQ Collaboration, H.~L. Lai \emph{et al.},
Eur.\ Phys.\ J. C \textbf{12}, 375 (2000).

\bibitem{QCDCorrections}
R.~Barbieri, R.~Gatto, R.~K\"ogerler, and Z.~Kunszt,
Phys.\ Lett.\ B \textbf{57}, 455 (1975);
R.~Barbieri, M.~Caffo, R.~Gatto, and E.~Remiddi,
Nucl.\ Phys.\ \textbf{B192}, 61 (1981).

\bibitem{QPM}
W.~Lucha, F.~F.~Schoberl, and D.~Gromes,
Phys.\ Rept.\ \textbf{200}, 127 (1991);
E.~J.~Eichten and C.~Quigg,
Phys.\ Rev.\ D \textbf{52}, 1726 (1995).

\bibitem{kkms}
M.~Klasen, B.~A.~Kniehl, L.~N.~Mihaila, and M.~Steinhauser,
Nucl.\ Phys.\ \textbf{B609}, 518 (2001);
Phys.\ Rev.\ Lett.\ \textbf{89}, 032001 (2002);
Nucl.\ Phys.\ \textbf{B713}, 487 (2005);
Phys.\ Rev.\ D \textbf{71}, 014016 (2005).

\end{thebibliography}
\end{document}